\documentclass[12pt]{article}

\usepackage{hyperref}
\usepackage{float}
\usepackage[utf8]{inputenc}
\usepackage{fullpage}
\usepackage{fullpage, graphicx,psfrag}
\usepackage{hyperref}
\usepackage[normalem]{ulem}
\usepackage[ruled,vlined]{algorithm2e}
\usepackage{color}
\usepackage{natbib}
\usepackage{amsmath}
\DeclareMathOperator*{\argmax}{arg\,max}

\newcommand{\INPUT}{\item[\textbf{Input:}]}
\newcommand{\OUTPUT}{\item[\textbf{Output:}]}
\def\boxit#1{\vbox{\hrule\hbox{\vrule\kern6pt\vbox{\kern6pt#1\kern6pt}\kern6pt\vrule}\hrule}}

\usepackage{multirow}
\usepackage{booktabs}
\usepackage{color}
\usepackage{url}
\usepackage{setspace}
\usepackage{amsmath,amsfonts,amssymb,amsthm}
         \usepackage{algorithmic}   
\usepackage{hyperref}
\usepackage{mathtools}
\usepackage{subcaption}

\newtheorem{remark}{Remark}

\usepackage{xcolor} 
\usepackage{cancel}

\allowdisplaybreaks

\usepackage{scalerel,stackengine}
\stackMath
\newcommand\reallywidehat[1]{%
\savestack{\tmpbox}{\stretchto{%
  \scaleto{%
    \scalerel*[\widthof{\ensuremath{#1}}]{\kern-.6pt\bigwedge\kern-.6pt}%
    {\rule[-\textheight/2]{1ex}{\textheight}}
  }{\textheight}%
}{0.5ex}}%
\stackon[1pt]{#1}{\tmpbox}%
}

\title{Nonparametric Linear Discriminant Analysis for High Dimensional Matrix-Valued Data}

\author{Seungyeon Oh \thanks{Department of Statistics, Sookmyung Women’s University, Seoul, Korea, \texttt{thisissyoh@gmail.com}}, Seongoh Park \thanks{ 
School of Mathematics, Statistics and Data Science at Sungshin Women's University, Seoul, Korea, \texttt{spark6@sungshin.ac.kr }},
~and Hoyoung Park \thanks{Department of Statistics, Research Institute of Natural Science, Sookmyung Women’s University, Seoul, Korea,  \texttt{hyparks@sookmyung.ac.kr}}}
\date{}

\begin{document}
\maketitle
	\begin{abstract}   
This paper addresses classification problems with matrix-valued data, 
{which commonly arise in applications such as neuroimaging and signal processing.} 
{Building on the assumption that the data from each class follows a matrix normal distribution, we propose a novel extension of Fisher's Linear Discriminant Analysis (LDA) tailored for matrix-valued observations. To effectively capture structural information while maintaining estimation flexibility, we adopt a nonparametric empirical Bayes framework based on Nonparametric Maximum Likelihood Estimation (NPMLE), applied to vectorized and scaled matrices.}
{The NPMLE method has been shown to provide robust, flexible, and accurate estimates for vector-valued data with 
{various structures in the mean vector or covariance matrix. By leveraging its strengths, our method is effectively generalized to the matrix setting, thereby improving classification performance.} Through extensive simulation studies and real data applications, including electroencephalography (EEG) and magnetic resonance imaging (MRI) analysis, we demonstrate that the proposed method tends to outperform} existing approaches across a variety of data structures.
 \bigskip 
 

\end{abstract}

\section{Introduction}

Matrix-valued data classification has become an essential challenge in modern data analysis, with applications spanning a variety of domains such as imaging studies, spatio-temporal data analysis, and multivariate time series. 
{Matrix-valued data inherently encodes richer structural information than traditional vector-valued data, capturing complex structures across rows and columns. For example, in signal datasets, it can effectively incorporate spatial and temporal dependencies that are essential for identifying meaningful patterns.} However, the classification of matrix-valued data often faces several key challenges. 
First, it is not straightforward to preserve the structural information inherent in matrix data when applying it to machine learning models. 
{In particular, traditional vectorization approaches, which simply flatten a matrix into a high-dimensional vector, 
{tend to ignore correlation structures among rows and columns. This often leads to a loss of essential information and results in suboptimal model performance.} 
Second, the curse of dimensionality and the accompanying computational burden become increasingly severe as the matrix dimensions grow, further complicating the classification task.} 


In recent years, significant progress has been made in developing methods to effectively handle matrix-valued data. 
{Most existing methodologies adopt the Fisher's Linear Discriminant Analysis (LDA), and penalized estimation strategies to overcome the high dimensionality of matrix-valued data.} 
{
{Among them, the penalized matrix linear discriminant analysis (PMLDA) proposed by \citet{molstad2019penalized} is a representative example. PMLDA assumes that the precision matrix of the vectorized representation admits a Kronecker product decomposition. It then jointly estimates the group mean vectors and the precision matrix required for LDA, while imposing sparsity-inducing penalties on both.}
{Around the same time, \cite{pan2019covariate} introduced the covariate-adjusted tensor classification in high dimensions (CATCH). The CATCH is also based on LDA, and it jointly utilizes both a low-dimensional covariate vector and a tensor predictor to estimate the response variable.} It assumes that the parameters in the LDA are sparse to address the high dimensionality.
Another notable approach is matrix linear discriminant analysis (MLDA) by \cite{hu_matrix_2020}. 
It starts by reformulating the LDA as a specific regression problem. 
{Subsequently, it utilizes the nuclear norm penalty to control the rank of the coefficient matrix of the regression model, enabling the extraction of informative structures within the matrix-valued data.} It also employs accelerated proximal gradient methods to solve the optimization problem that is not differentiable but convex. Lastly, an important advancement is sparse and reduced-rank linear discriminant analysis (Sr-LDA) by \cite{wang_sr_2024}. 
{By incorporating both nuclear norm and $\ell_1$ norm penalties to estimate the coefficient matrix in the LDA, the method successfully reflects both sparsity and low-rank structures. This dual-penalty approach significantly improves interpretability and classification accuracy, thereby offering a robust solution for high-dimensional matrix-valued data.}
Beyond penalized LDA formulations, generative modeling under heavy tails has also been explored. For example, \citet{thompson_classification_2020} develop classification rules under the matrix-variate $t$ distribution, which can offer robustness to outliers and tail misspecification relative to matrix-normal assumptions. In a different direction, \citet{zhang_study_2025} study overlap in matrix-variate data and formalize overlap measures with applications to discriminant analysis, providing classifier-agnostic diagnostics of intrinsic problem difficulty.
Despite these advances across penalized, generative, and diagnostic perspectives, many methods in the broader literature still rely on structural assumptions (e.g., sparsity, low rank, separability) that may not align with certain real-world datasets; when the true data-generating mechanism deviates from these assumptions, the performance of these methods can deteriorate substantially.


{
In parallel, extensive research has been conducted to improve LDA for vector-valued data. Specifically, the classical LDA relies on the sample mean and sample covariance matrix, which can be considered naive estimators in high-dimensional low sample size settings, as they suffer from rank deficiency and estimation instability. In particular, \citet{greenshtein_application_2009}, \citet{Park_2022}, and \citet{MVA_2024} demonstrate that nonparametric empirical Bayes (NPEB) methods yield improved classification performance by estimating group means more accurately.
Building on this insight, our study aims to extend the NPEB-based modifications of LDA, proposed for vector-valued data, to accommodate matrix-valued data. More specifically, we refine the matrix version of LDA by scaling the vectorized data using suitable precision matrix estimators, and then estimating the scaled mean parameters through NPEB, in line with the rationale of \cite{Park_2022}. 
{Unlike existing methods that depend heavily on penalization strategies to regularize high-dimensional structures, our method adapts automatically to a wide range of data-generating mechanisms without the need for manual tuning or assumption-specific modeling. 
Notably, it maintains robustness across diverse scenarios, including settings that violate the structural assumptions underlying existing approaches. These results highlight the practical utility and broad applicability of our method for high-dimensional matrix-valued data classification. }

{The forthcoming sections are organized as follows. In Section \ref{sec:Method}, we address the parameter estimation techniques for LDA via NPMLE, and subsequently propose a novel and flexible classification rule for matrix-valued data. Section \ref{sec:sim} and Section \ref{sec:da} cover extensive simulation studies and real-world case analyses, including applications to electroencephalography (EEG) and magnetic resonance imaging (MRI) dataset. Lastly, Section \ref{sec:conclusion} summarizes our key contributions and suggests directions for future research.}


\section{Methodology} \label{sec:Method}
This section introduces the high-dimensional parameter estimation method used to design a binary classifier for matrix-valued data. 
\subsection{
{Problem setup}}
{We begin by assuming that a feature matrix $\mathbf{X} \in \mathbb{R}^{p \times q}$ is drawn from a matrix normal distribution. Specifically, given the class label $Y \in \{1, 2\}$, we assume}  
\begin{align}
   \mathbf{X} \mid Y = k, \mathbf{M}_k, \mathbf{U}, \mathbf{V} &\sim \text{MN}(\mathbf{M}_k, \mathbf{U}, \mathbf{V}), \label{eq:dist1}
\end{align}
{where $\mathbf{M}_k \in \mathbb{R}^{p \times q}$ denotes the mean matrix of group $k$, $\mathbf{U} \in \mathbb{R}^{p \times p}$, and $\mathbf{V} \in \mathbb{R}^{q \times q}$ denote covariances among the rows and columns of $\mathbf{X}$.}
{The matrix normal distribution in Equation \eqref{eq:dist1} can also be represented as the following multivariate normal distribution:
\begin{align}
    \operatorname{vec}(\mathbf{X}) \mid Y = k, \mathbf{M}_k, \mathbf{U}, \mathbf{V} &\sim N\left( \operatorname{vec}(\mathbf{M}_k), \mathbf{V} \otimes \mathbf{U} \right), \label{eq:dist2}
\end{align}
where $\operatorname{vec(\cdot)}$ represents a vectorization operator, and $\otimes$ is the Kronecker product.} For simplicity of notation, we denote $\mathbf{V} \otimes \mathbf{U}$ by $\mathbf{\Sigma}$ throughout Section \ref{sec:Method}.
Under this model, the optimal Bayes classification rule can be derived by observing that the matrix normal model in \eqref{eq:dist1} is equivalent to the multivariate Normal formulation in \eqref{eq:dist2}. This leads to the following discriminant function:
\begin{align}
            \delta\left(\mathbf{X}^{\text{New}}\right) &=   \left( \operatorname{vec}\left(\mathbf{X}^{\text{New}}\right) - \frac{\operatorname{vec}(\mathbf{M}_1) + \operatorname{vec}(\mathbf{M}_2)}{2} \right)^\top 
            \left( \mathbf{V} \otimes \mathbf{U} \right) ^{-1} 
           \left( \operatorname{vec}(\mathbf{M}_1) - \operatorname{vec}(\mathbf{M}_2) \right) \nonumber \\
            &= \left(\mathbf{\Sigma}^{-\frac{1}{2}} \operatorname{vec}\left(\mathbf{X}^{\text{New}}\right) - \frac{\mathbf{\Sigma}^{-\frac{1}{2}}  \left( \operatorname{vec}(\mathbf{M}_1) + \operatorname{vec}(\mathbf{M}_2) \right)}{2} \right)^\top 
            \mathbf{\Sigma}^{-\frac{1}{2}}
            \left( \operatorname{vec}(\mathbf{M}_1) - \operatorname{vec}(\mathbf{M}_2) \right).  \label{eq:cr}
\end{align}
\noindent        
Then, using \eqref{eq:cr}, we can classify $Y$ as follows:
\begin{eqnarray}
    \widehat{Y}^{\text{New}} &=& 
\begin{cases} 
1 & \text{for } \delta(\mathbf{X}^{\text{New}}) \geq \log\left(\pi_2 / \pi_1\right), \\
2 & \text{for } \delta(\mathbf{X}^{\text{New}}) < \log\left(\pi_2 / \pi_1\right).
\label{eq:cr2} \end{cases}
\end{eqnarray}
{
However, in the absence of information about the mean matrix $\mathbf{M}_k$ for the two groups and the covariance matrices $\mathbf{U}$ and $\mathbf{V}$, the classification rule in Equation \eqref{eq:cr} and \eqref{eq:cr2} cannot be applied in practice. 
As emphasized in \cite{Park_2022}, approximating the discriminant function \eqref{eq:cr} by substituting the unknown parameters with suitable estimators can be a practical and effective choice.
To this end, inspired by the theoretical contributions in \cite{Park_2022}, we obtain an estimator of the $\mathbf{\Sigma}^{-\frac{1}{2}}$, subsequently estimating the scaled mean vectors $\mathbf{\Sigma}^{-\frac{1}{2}} \operatorname{vec}(\mathbf{M}_1)$ and $\mathbf{\Sigma}^{-\frac{1}{2}} \operatorname{vec}(\mathbf{M}_2)$ using nonparametric maximum likelihood estimation (NPMLE)} approach within an empirical Bayes framework. This strategy allows for a more flexible approximation to the discriminant function, while avoiding strict structural assumption on the scaled mean. As a result, it has the potential to enhance both classification accuracy and applicability to a broad range of matrix-valued data scenarios.

\subsection{Parameter estimation}
\label{eqn:subsec:parameter_estimation}
{In this section, we introduce the procedure for estimating the scaled group-specific mean vectors using the nonparametric maximum likelihood estimation (NPMLE) method, assuming that the common covariance structure of the two matrix normal distributions is known. This procedure is later extended to the case where the covariance matrix is estimated from the data.} We aim to directly estimate the scaled mean vectors in the discriminant function \eqref{eq:cr}, denoted by $\mathbf{\Sigma}^{-\frac{1}{2}} \operatorname{vec}(\mathbf{M}_1)$ and $\mathbf{\Sigma}^{-\frac{1}{2}} \operatorname{vec}(\mathbf{M}_2)$. For this purpose, it is necessary to investigate the distribution of the data scaled by $\mathbf{\Sigma}^{-\frac{1}{2}}$.

Let the $i$-th observed matrix-valued data in the $k$-th group be $\left\{\left(Y_i^{(k)}, \mathbf{X}_i^{(k)}\right)\right\}_{i=1}^{n_k}$, where $\mathbf{X}_i^{(k)} \in \mathbb{R}^{p \times q}$ is a feature matrix and $Y_i^{(k)}$ is its associated label indicating group $k$. Let $n_k$ denote the number of samples belonging to the $k$-th group. Then we assume that $\mathbf{X}_i^{(k)}$ follows a matrix normal distribution. Specifically, we can state the following:
{\begin{align}
   \mathbf{X}_i^{(k)} \mid Y_i^{(k)}, \mathbf{M}_k, \mathbf{U}, \mathbf{V} &\sim \text{MN}(\mathbf{M}_k, \mathbf{U}, \mathbf{V}), \quad i \in \{1, \dots, n_k\}, \quad k \in \{1, 2\},
   \label{eq:dist_x_i_prev}
\end{align}
which is equivalent to stating that $\operatorname{vec}\left(\mathbf{X}_i^{(k)}\right)$ follows a multivariate Normal distribution:
\begin{align}
\operatorname{vec}\left(\mathbf{X}_i^{(k)}\right) \mid Y_i^{(k)}, \mathbf{M}_k, \mathbf{\Sigma} &\sim N\left( \operatorname{vec}(\mathbf{M}_k), \mathbf{\Sigma} \right), \quad i \in \{1, \dots, n_k\}, \quad  k \in \{1, 2\}, \label{eq:dist_x_i}
\end{align}
where $\mathbf{\Sigma}$ denotes $\mathbf{V} \otimes \mathbf{U}$.} 

{For each group, we aim to decorrelate the observed data $\operatorname{vec}\left(\mathbf{X}_i^{(k)}\right)$ and rescale its variance to unity. To achieve this, we define the scaled vectorized observations as} 
$\mathbf{z}_i^{(k)} = \left(z_{i,1}^{(k)}, \cdots, z_{i, pq}^{(k)}\right)^{\top} 
    := \mathbf{\Sigma}^{-\frac{1}{2}} \operatorname{vec}\left(\mathbf{X}_i^{(k)}\right), \
    i \in \{1, \dots, n_k\}, \  k \in \{1, 2\}$,
then, from \eqref{eq:dist_x_i}, we can derive the conditional distribution of $\mathbf{z}_i^{(k)}$ given $Y_i^{(k)}$ as follows:
\begin{eqnarray}
    \mathbf{z}_i^{(k)} & \mid Y_i^{(k)}, \mathbf{M}_k, \mathbf{\Sigma} 
     \sim N \left(\mathbf{\Sigma}^{-\frac{1}{2}} \operatorname{vec}(\mathbf{M}_k), \mathbf{I}\right). \label{eq:scaled_dist}
\end{eqnarray}
}

{To obtain summary statistics for the scaled mean vector in Equation \eqref{eq:scaled_dist},} 
{we define the $j$-th sample mean of the scaled data $\bar{z}_{.j} ^{(k)}$ for $j \in \{1, \dots, pq\}$ as:
$$
\bar{z}_{.j}^{(k)} := \frac{1}{n_k} \sum_{i=1}^{n_k} z_{i,j}^{(k)}, \quad k \in \{1, 2\},
$$
and denote $\mathbf{\Sigma}^{-\frac{1}{2}} \operatorname{vec}(\mathbf{M}_k)$ by $ \boldsymbol{\mathcal{M}}_k = \left( m_1^{(k)}, \cdots, m_{pq}^{(k)} \right)^ \top$ to avoid notational complexity. Then, to simulataneously estimate the components of the scaled mean vector based on these summary statistics, we apply the Bayesian hierarchical model to each obtained $\bar{z}_{.j}^{(k)}$, $j \in \{1, \dots, pq\}$ as:
\begin{align}
    \bar{z}_{.j}^{(k)} \mid \boldsymbol{\mathcal{M}}_k & \overset{ind}{\sim}N\left(m_j^{(k)}, \frac{1}{n_k}\right), \quad k \in \{1, 2\}, \label{eq:model1}\\
    m_j^{(k)} & \overset{iid}{\sim}G^{(k)} \in \mathcal{G}, \quad k \in \{1, 2\}, \label{eq:model2}
\end{align}
}
where $\mathcal{G}$ contains all possible distributions with the support of $(-\infty, \infty)$, and $G^{(k)}$ is an unknown distribution for $m_j^{(k)}, \ j \in \{1, \dots, pq\}$. Here, our objective is to determine the distribution $G^{(k)}$ which best captures the characteristics of the scaled data through non-parametric maximum likelihood estimation (NPMLE). 
Let $f(\cdot)$ be the marginal probability density function of $\bar{z}_{.j}^{(k)}$, and let $g(\cdot \mid m)$ be the conditional probability density function of $\bar{z}_{.j}^{(k)}$ given $m$. Then, it is natural to consider the distribution which maximizes the likelihood of observing the $\bar{z}_{.j}^{(k)}, \ j \in \{1, \dots, pq\}$ as the distribution $G^{(k)}$. Looked at from that point of view, the optimization process can be expressed as Equation \eqref{eq:opt}. 
{To solve the infinite-dimensional optimization problem, we employ the efficient algorithms, such as \cite{lindsay_mixture_1995},\cite{koenker_convex_2014}, \cite{dicker_high_2016}, and \cite{feng_approximate_2016}. These algorithms reformulate the infinite-dimensional problem as a finite-dimensional one by approximating $G^{(k)}$ by a finite mixture model.} Then, the estimator of the distribution $G^{(k)}$ can be expressed as 

\begin{align}
    \widehat{G}^{(k)} & = \argmax_{G^{(k)} \in \mathcal{G}^{(k)}} \sum_{j=1}^{p q} \log f\left(\bar{z}_{.j}^{(k)}\right) \nonumber\\
&= \argmax_{G^{(k)} \in \mathcal{G}^{(k)}} \sum_{j=1}^{pq} \log \left\{ \int \frac{\sqrt{n_k}}{\sqrt{2\pi}} \exp \left\{ - \frac{n_k\left(\bar{z}_{.j}^{(k)} - m\right)^2}{2} \right\} dG^{(k)}(m) \right\}. \label{eq:opt} 
\end{align}
Using the solution for $\widehat{G}^{(k)}$ in \eqref{eq:opt}, we obtain the estimator for the $j$-th element of the scaled mean vector, denoted by $\widehat{m}_j^{(k)}$, as follows:
 \begin{align}
            \displaystyle \widehat{m}_j^{(k)} &= {E}_{\widehat{G}}\left[m|\bar{z}_{.j}^{(k)}\right]
            \nonumber\\
            &= \displaystyle \frac{\int m \, g\left(\bar{z}_{.j}^{(k)}|m\right) \, d\ \widehat{G}^{(k)}(m)}{\int g\left(\bar{z}_{.j}^{(k)}|m\right) \, d\widehat{G}^{(k)}(m)}  = \frac{\sum_{ \ell=1}^{L} \nu_{ \ell} \widehat{\omega}_{\ell} g\left(\bar{z}_{.j}^{(k)}|\nu_{\ell}\right)}{\sum_{{\ell}=1}^{L} \widehat{\omega}_{\ell} g\left(\bar{z}_{.j}^{(k)}|\nu_{\ell}\right) }. \label{eq:est_smean}
\end{align}
where $L$ is the number of grid points of the finite mixture model for $m_j^{(k)}$, $\nu_{\ell}$ is the value of the ${\ell}$-th grid point, and $\widehat{\omega}_{\ell}$ is the weight corresponding to $\nu_{\ell}$.

\subsection{Proposed classification rule}\label{sec:cr}

 {To make the classification rule \eqref{eq:cr} and \eqref{eq:cr2} practical, we need to approximate the rule by estimating the unknown parameters, such as $\mathbf{\Sigma}^{-\frac{1}{2}}$ and $\mathbf{\Sigma}^{-\frac{1}{2}} \operatorname{vec}(\mathbf{M}_k)$. Accordingly, we consider combining the parameter estimation techniques for the scaled mean vectors introduced in Section  \ref{eqn:subsec:parameter_estimation} with specialized covariance or precision estimation methods for matrix-valued data. However, in the matrix normal setting, the covariances along the row and column directions, $\mathbf{U}$ and $\mathbf{V}$, are identifiable only up to a scaling constant due to  he Kronecker product structure.  We recognize the problem and refer to existing methods such as the graph estimation for matrix-variate normal instances (GEMINI) proposed by \cite{shuheng_2014}, and \cite{Zhang_2023}. Among them, we apply the GEMINI method, which determines the scale based on weight matrices and the average squared Frobenius norm of the samples. Indeed, we adopt this approach in both the simulation study in Section \ref{sec:sim} and the real data analysis in Section \ref{sec:da}. } However, readers are encouraged to consider alternative estimators for $\mathbf{\Sigma}^{-\frac{1}{2}}$ according to their specific preferences or application contexts. As a result, the core idea of this paper is formulated through the following algorithm. 

\begin{algorithm}
\caption{The classification rule of NPMLDA}\label{alg:NPMLDA_alg}
\begin{algorithmic}[1]
\INPUT The observed data : $\{(Y_i^{(k)}, \mathbf{X}_i^{(k)})\}_{i=1}^{n_k}$, $\mathbf{X}_i^{(k)} \in \mathbb{R}^{p \times q} $, $Y_i^{(k)} \in \{1,2\}, \ k \in \{1,2\}$.
\OUTPUT
\STATE Using the covariance estimation methods briefly mentioned above, obtain an estimate of $\mathbf{\Sigma}^{-\frac{1}{2}} = \mathbf{V}^{-\frac{1}{2}} \otimes \mathbf{U}^{-\frac{1}{2}}$.
\STATE Vectorize the observed data, and rescale them using $\widehat{\mathbf{\Sigma}}^{-\frac{1}{2}}$ : $\mathbf{z}_i^{(k)} = \left(z_{i,1}^{(k)}, \cdots, z_{i, pq}^{(k)}\right)^{\top} 
    := \widehat{\mathbf{\Sigma}}^{-\frac{1}{2}}\operatorname{vec}\left(\mathbf{X}_i^{(k)}\right), \
    \text{for all } i \text{ and } k$. 
\STATE For each group $k$, compute the $j$-th sample mean $\bar{z}_{.j}^{(k)} := \frac{1}{n_k} \sum_{i=1}^{n_k} z_{i,j}^{(k)}$ for $j \in \{1, \dots pq\}$, and apply the Bayesian hierarchical model \eqref{eq:model1} and \eqref{eq:model2} to the sample means. 
\STATE Estimate the distribution $G^{(k)}$ for elements of the scaled mean vector in \eqref{eq:model2} via the NPMLE framework.
\STATE Based on the estimated distribution $\widehat{G}^{(k)}$, obtain the Bayes estimator \eqref{eq:est_smean} for the scaled mean vector $\widehat{\boldsymbol{\mathcal{M}}}_k := \reallywidehat{\boldsymbol{\Sigma}^{-\frac{1}{2}} \operatorname{vec}(\mathbf{M}_k)}$.
\STATE Classify the new observation $\mathbf{X}^{\text{new}}$ according to the proposed classification rule below: 
\begin{align}
        \widehat{\delta} (\mathbf{X}^{\text{new}}) 
        &= \left( \widehat{\boldsymbol{\Sigma}}^{-\frac{1}{2}} \operatorname{vec}(\mathbf{X}^{\text{new}}) 
        - \frac{\widehat{\boldsymbol{\mathcal{M}}}_1 + \widehat{\boldsymbol{\mathcal{M}}}_2}{2} \right)^\top 
        \left( \widehat{\boldsymbol{\mathcal{M}}}_1 - \widehat{\boldsymbol{\mathcal{M}}}_2 \right) - \log\left({\frac{\widehat{\pi}_2}{\widehat{\pi}_1}}\right), \\
            \widehat{Y}^{\text{New}} &= 
            \begin{cases} 
            1 & \text{for } \widehat{\delta}(\mathbf{X}^{\text{New}}) \geq 0, \nonumber \\
            2 & \text{for } \widehat{\delta}(\mathbf{X}^{\text{New}}) < 0, \nonumber
            \end{cases} 
\end{align}

{where $\widehat{\pi}_k$ is the proportion of observations belonging to $k$-th group among the all observations.}
\end{algorithmic}
\end{algorithm}

\begin{remark}\label{remark:multiclass}
\textnormal{
{The proposed classification rule can be naturally extended to the multi-class setting $(K \geq 3)$ as follows. In this context, we consider the classification task among $K$ distinct groups, where each observation $\mathbf{X} \in \mathbb{R}^{p \times q}$ is assumed to follow a matrix normal distribution with group-specific mean matrix and a common covariance structure. To classify a new observation $\mathbf{X}^{\text{New}}$, we compute the posterior probability that it belongs to each class $k \in \{1, \dots, K\}$, and assign it to the class with the highest posterior. Adopting the scaled mean vector representation $\widehat{\boldsymbol{\mathcal{M}}}_k := \reallywidehat{\boldsymbol{\Sigma}^{-\frac{1}{2}} \operatorname{vec}(\mathbf{M}_k)}$, the classification rule takes the following form:}
}
\begin{align}
\widehat{Y}^{\text{\textnormal{New}}} &= \argmax_{k \in \{1, \dots, K\}} \ P(Y^{\text{\textnormal{New}}}=k |\mathbf{X^{\text{\textnormal{New}}}}) \nonumber \\
&= \argmax_{k \in \{1, \dots, K\}} \ \log{\widehat{\pi}_k} -\frac{1}{2} \left( \widehat{\boldsymbol{\Sigma}}^{-\frac{1}{2}} \, \mathrm{vec}(\mathbf{X}^{\text{\textnormal{New}}}) - \widehat{\boldsymbol{\mathcal{M}}}_k  \right)^\top 
        \left( \widehat{\boldsymbol{\Sigma}}^{-\frac{1}{2}} \mathrm{vec}(\mathbf{X}^{\text{\textnormal{New}}}) - \widehat{\boldsymbol{\mathcal{M}}}_k  \right)  
    \label{eq:multiclass}
\end{align}
\textnormal{
{The feasibility of this multiclass extension is supported by the parameter estimation techniques introduced in Section~\ref{eqn:subsec:parameter_estimation}, where the scaled mean vectors $\widehat{\boldsymbol{\mathcal{M}}}_k$ are estimated in a nonparametric fashion using the NPMLE approach. 
{Consequently, the proposed method retains its flexibility even in the multiclass setting, avoiding additional structural assumptions beyond the normality assumption.}}}
\end{remark}

\section{Simulation study} \label{sec:sim}
This section presents a series of experiments to assess the performance of our approach. 
{We compare our classifier against several state-of-the-art that have demonstrated remarkable performance in matrix-valued data classification.}
Specifically, we include the penalized matrix linear discriminant analysis (PMLDA) by \cite{molstad2019penalized}, the covariate-adjusted tensor classification in high dimensions (CATCH)  by \cite{pan2019covariate}, the matrix linear discriminant analysis (MLDA) by \cite{hu_matrix_2020}, and the sparse and reduced rank linear discriminant analysis (Sr-LDA) by \cite{wang_sr_2024}. 
{In addition to these competing methods, we also consider an oracle classifier that utilizes the true model parameters in the discriminant function \eqref{eq:cr}, serving as a performance benchmark.}
{To highlight the robustness and adaptability of our method, we evaluate all classifiers under a variety of matrix structures. 
} This contrasts with prior studies that typically assume specific structural forms, allowing us to demonstrate the broader applicability of our approach.

Given the label variable $Y_i^{(k)}$, the feature matrix $\mathbf{X}_i^{(k)}$ is generated from  $\text{MN}(\mathbf{M}_k, \mathbf{U}, \mathbf{V})$, where $\mathbf{M}_k \in \mathbb{R}^{p \times q}, \mathbf{U} \in \mathbb{R}^{p \times p}$, and $\mathbf{V} \in \mathbb{R}^{q \times q}$. 
{We define the coefficient matrix of the discriminant function in \eqref{eq:cr}, derived from Bayes' rule, as $\mathbf{B} = \mathbf{U}^{-1} (\mathbf{M}_1 - \mathbf{M}_2) \mathbf{V}^{-1}$.}
In most studies, the structure of $\mathbf{B}$ is often assumed to be sparse, implying that most elements of $\mathbf{B}$ are exactly zero. Conversely, our simulation study explores both scenarios: one where the structure of $\mathbf{B}$ is sparse and another where $\mathbf{B}$ is dense, indicating that none of its elements are zero. Additionally, inspired by \cite{pan2019covariate} and \cite{wang_sr_2024}, we address three models related to the covariance matrices $\mathbf{U}$ and $\mathbf{V}$. The details are as follows : 
\begin{figure}[H]
    \centering
    \includegraphics[width=0.7\linewidth]{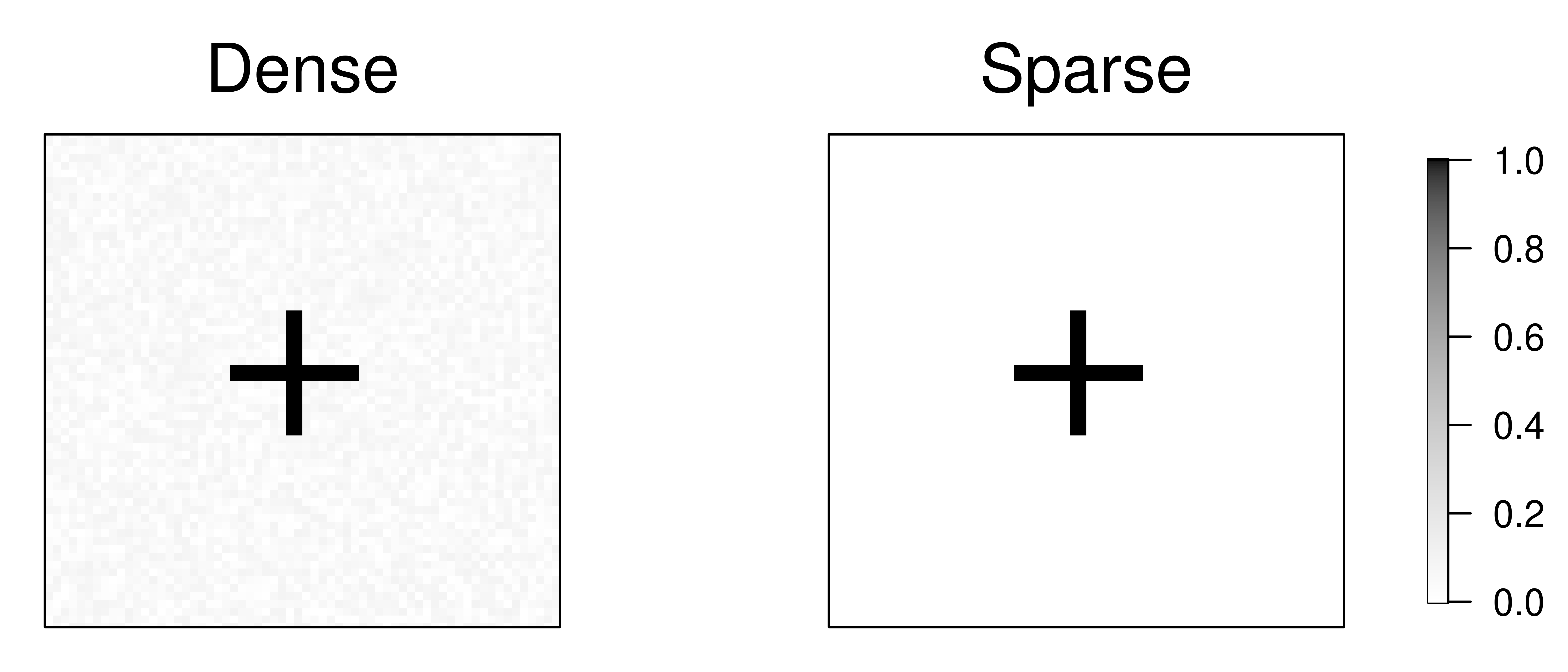}
    \caption{Examples of $\mathbf{B}$ at $\theta = 1$ : dense (left) and sparse (right) settings.}
    \label{fig:B_examples}
\end{figure}

\begin{enumerate}
    \item A feature matrix is a two-dimensional matrix with $p=64$ rows and $q=64$ columns.
    \item We generate 300 training data and 300 test data for each group; that is, $n_1 =n_2 = 300$.
    \item The mean matrix for group 1 is defined as $\mathbf{M}_1 := 0_{p \times q}$, and for group 2, as $\mathbf{M}_2 := - \mathbf{U}\mathbf{B}\mathbf{V}$.
    \item 
    {We consider both scenarios in which $\mathbf{B}$ is sparse, and in which it is dense. First, each sparse $\mathbf{B}$ consists of a uniformly white background and a cross pattern. In contrast, each dense $\mathbf{B}$ consists of a noisy background and a cross pattern. As a background, the white color denotes value 0, and each value of noise is generated from $\text{Unif}(0,0.1)$. Additionally, we consider $\theta$, which represents the values of the black color, to be $(0.9,1.0,1.1)$ to enhance the variety in our experiments. The Figure \ref{fig:B_examples} illustrates the examples of $\mathbf{B}$.}
    \item We consider three models related to covariance matrices $\mathbf{U}$ and $\mathbf{V}$ :
    \begin{itemize}
        \item (Model 1) 
        \item [] Model 1 represents a doubly independent case where the both the $\mathbf{U}$ and $\mathbf{V}$ are diagonal matrices. Specifically, we set $\mathbf{U} = 1/5 \cdot \mathbf{I}_{p \times p}$, and $\mathbf{V} = 1/5 \cdot \mathbf{I}_{q \times q}$.
        \item (Model 2) 
        \item [] Model 2 represents a case where the rows of the data are independent of each other, while the columns are not. Here, we set $\mathbf{U} = 1/5 \cdot \mathbf{I}_{p \times p}$. For $\mathbf{V}$, we consider the autoregressive model $\text{AR}(1)$ with autocorrelation coefficient $\rho=0.5$. Additionally, after scaling, we set $\mathbf{V}_{ij}= 1/5 \cdot 0.5^{|i-j|}$.
        \item (Model 3) 
        \item [] Model 3 represents a doubly dependent case where $\mathbf{U}_{ij}=1/5 \cdot 0.25^{|i-j|}, \ \mathbf{V}_{ij}= 1/5 \cdot 0.25^{|i-j|}$.
    \end{itemize}
\end{enumerate}
{We evaluate the models by generating 200 independent datasets using the same procedure, computing the misclassification rate for each. The elased computing times are presented in Table \ref{tab:sim_time}, and were measured in R version 4.5.1 on an Ubuntu 22.04.5 LTS workstation with an Intel(R) Xeon(R) Gold 5220R CPU @ 2.20GHz.}

\begin{table}[H]
\centering
\begin{tabular}{c|c|ccccc}
\toprule
\multirow{2}{*}{$\mathbf{B}$}      & \multirow{2}{*}{\textbf{Model}} & \multicolumn{5}{c}{\textbf{Method}}             \\ \cline{3-7} 
                        &                        & \textbf{NPMLDA} & \textbf{MLDA} & \textbf{PMLDA} & \textbf{CATCH} & \textbf{Sr-LDA} \\ \hline
\multirow{3}{*}{Dense}  & 1                      & 272    & 483  & 58    & 21    & 1446   \\
                        & 2                      & 274    & 525  & 60    & 32    & 1009    \\
                        & 3                      & 276    & 497  & 66    & 28    & 1018    \\ \hline
\multirow{3}{*}{Sparse} & 1                      & 270    & 444  & 58    & 21    & 1782     \\
                        & 2                      & 271    & 519  & 59    & 30    & 1027     \\
                        & 3                      & 272    & 468  & 67    & 29    & 1212     \\ \bottomrule
\end{tabular}
\caption{Elapsed computing times (sec) of classifiers for a single iteration.}
\label{tab:sim_time}
\end{table}


\begin{figure}[H]
    \centering
    \includegraphics[width=0.85\linewidth]{figures/simulation2_box/dense_model1.png}
    \caption{Misclassification rates in Model 1 under the dense $\mathbf{B}$ setting.}
    \label{fig:dense_M1}
\end{figure}

\begin{figure}[H]
    \centering
    \includegraphics[width=0.85\linewidth]{figures/simulation2_box/sparse_model1.png}
    \caption{Misclassification rates in Model 1 under the sparse $\mathbf{B}$ setting.}
    \label{fig:sparse_M1}
\end{figure} 

{First, Figures \ref{fig:dense_M1} and \ref{fig:sparse_M1} show misclassification rates of the classifiers under Model 1 (the doubly independent case) where the former represents the dense $\mathbf{B}$ setting, and the latter represents the sparse $\mathbf{B}$ setting. In these settings, regardless of the sparsity of $\mathbf{B}$, our NPMLDA exhibits strong empirical performance, closely approaching that of the oracle classifier, which ensures the theoretical optimum. Sr-LDA tends to follow NPMLDA in terms of performance, while the remaining classifiers occasionally fail to even approach NPMLDA, let alone the oracle classifier.}




\begin{figure}[H]
    \centering
    \includegraphics[width=0.85\linewidth]{figures/simulation2_box/dense_model2.png}
    \caption{Misclassification rates in Model 2 under the dense $\mathbf{B}$ setting.}
    \label{fig:dense_M2}
\end{figure}

\begin{figure}[H]
    \centering
    \includegraphics[width=0.85\linewidth]{figures/simulation2_box/sparse_model2.png}
    \caption{Misclassification rates in Model 2 under the sparse $\mathbf{B}$ setting.}
    \label{fig:sparse_M2}
\end{figure}

Second, Figures \ref{fig:dense_M2} and \ref{fig:sparse_M2} illustrate the performance of classifiers under Model 2 (independent rows and dependent columns case), with the former assuming dense $\mathbf{B}$, and the latter assuming sparse $\mathbf{B}$. 
{In Figure \ref{fig:dense_M2}, which corresponds to the dense $\mathbf{B}$ setting, NPMLDA shows the lowest misclassification rates among the classifiers, except for the oracle model. Meanwhile, in Figure \ref{fig:sparse_M2}, which corresponds to the sparse $\mathbf{B}$ setting, classifiers that are based on the sparsity assumption for $\mathbf{B}$ tend to perform well, unlike the results of the Model 1. In particular, Sr-LDA achieves relatively lower misclassification rates. However, we can observe that NPMLDA becomes increasingly comparable to it as $\theta$ closes to $1.1$.}





\begin{figure}[H]
    \centering
    \includegraphics[width=0.85\linewidth]{figures/simulation2_box/dense_model3.png}
    \caption{Misclassification rates in Model 3 under the dense $\mathbf{B}$ setting.}
    \label{fig:dense_M3}
\end{figure}

\begin{figure}[H]
    \centering
    \includegraphics[width=0.85\linewidth]{figures/simulation2_box/sparse_model3.png}
    \caption{Misclassification rates in Model 3 under the sparse $\mathbf{B}$ setting.}
    \label{fig:sparse_M3}
\end{figure}

Lastly, Figures \ref{fig:dense_M3} and \ref{fig:sparse_M3} present the results under Model 3 (the doubly dependent case), corresponding to the settings where $\mathbf{B}$ is dense and sparse, respectively. 
{Similar to the results of Model 1 and 2, NPMLDA generally outperforms the other non-oracle methods in the dense $\mathbf{B}$ settings. In the sparse $\mathbf{B}$ settings, our classifier and Sr-LDA appear to be indistinguishable in performance.}




So far, it has been considered that the majority of existing methods perform poorly under the dense $\mathbf{B}$ setting due to a mismatch between the data structure and the model assumptions. To verify whether the mismatch causes the performance degradation, we need to examine how each model estimates the coefficient matrix $\mathbf{B}$ of the discriminant function \eqref{eq:cr}. In this context, Figures \ref{fig:est_B_dense} and \ref{fig:est_B_sparse} present the average of the estimated $\mathbf{B}$ from 50 replications for each method. For clarity of visual interpretation, we adopt a larger cross pattern for the true $\mathbf{B}$ than in the previous experiments. First, under the dense $\mathbf{B}$ setting, the proposed NPMLDA provides estimates that are the closest to the original $\mathbf{B}$. However, Sr-LDA, PMLDA, and CATCH cannot reflect the information about the noisy backgrounds due to sparsity assumption on $\mathbf{B}$. Additionally, MLDA fails to clearly recover the cross pattern of the original $\mathbf{B}$. 
{Second, even under the sparse B setting, NPMLDA seems to most accurately recover the original $\mathbf{B}$. These results provide evidence for our simulation results while also implying that NPMLDA has the flexibility to the various structures since it estimates the mean parameters for LDA without relying on structural regularization.}

{In summary, NPMLDA demonstrates top-tier performance on average, and shows relatively stable misclassification rates across Monte Carlo replications. Conversely, the other classifiers tend to be relatively sensitive to the structures of $\mathbf{B}$ or the difficulty of the classification problem. In addition, through Figures \ref{fig:est_B_dense} and \ref{fig:est_B_sparse}, we demonstrate the reason for the poor performance of sparsity-based methods under the dense $\mathbf{B}$ setting, while clearly highlighting the superiority of NPMLDA across all settings in estimating $\mathbf{B}$. Therefore, these results empirically show that our NPMLDA can be applied without assuming that the mean matrix of the matrix-valued data are low-rank or sparse, while still achieving notable performance across a wide range of cases.}

{Additionally, we conduct extended experiments with smaller and larger cross patterns of $\mathbf{B}$ than those in the main experimental design, and the corresponding results can be found in the Supplementary Material.}
\begin{figure} [H]
        \centering
        \includegraphics[width=0.9\linewidth]{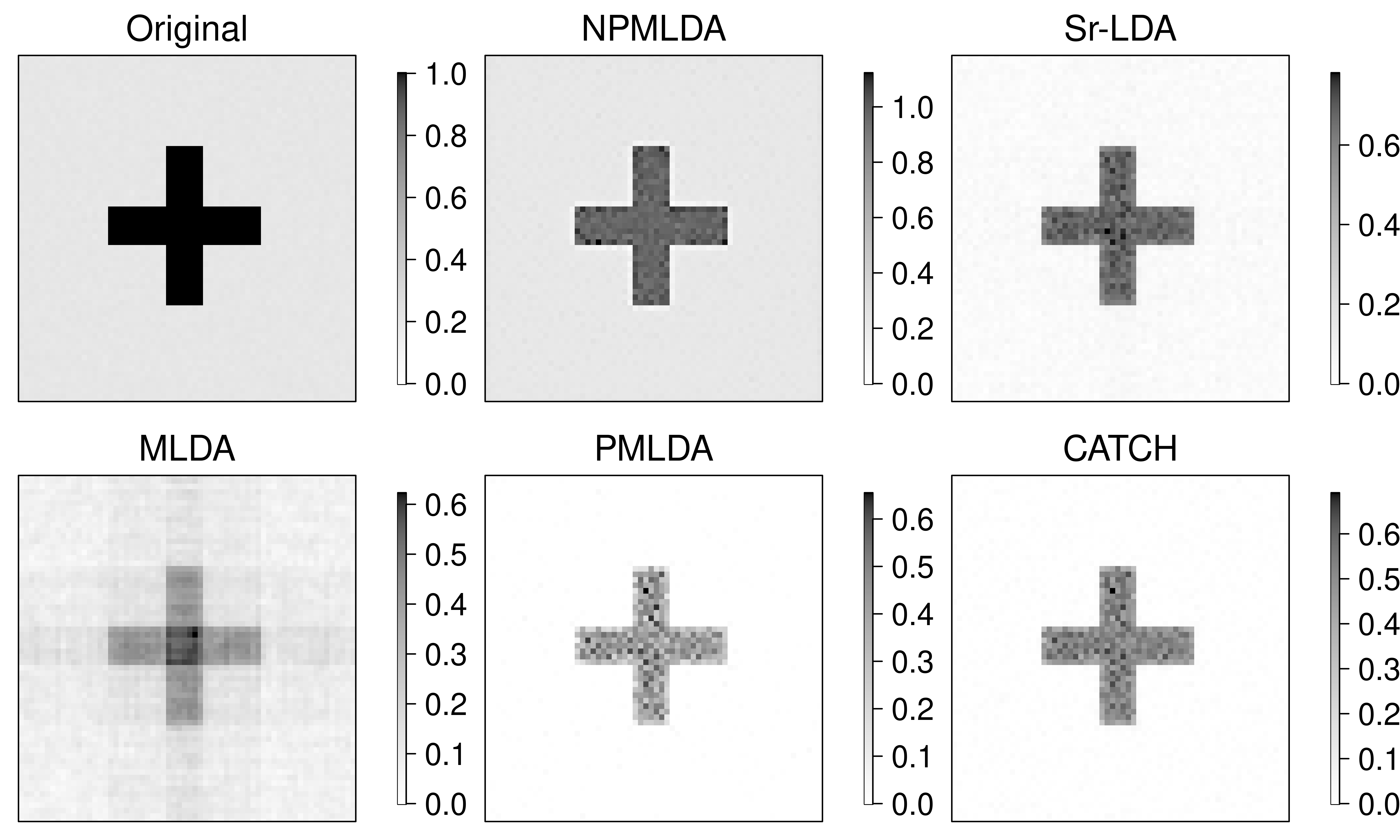}
        \caption{Averaged estimated $\mathbf{B}$ from each method under the dense $\mathbf{B}$ setting.}
        \label{fig:est_B_dense}
\end{figure}

\begin{figure} [H]
        \centering
        \includegraphics[width=0.9\linewidth]{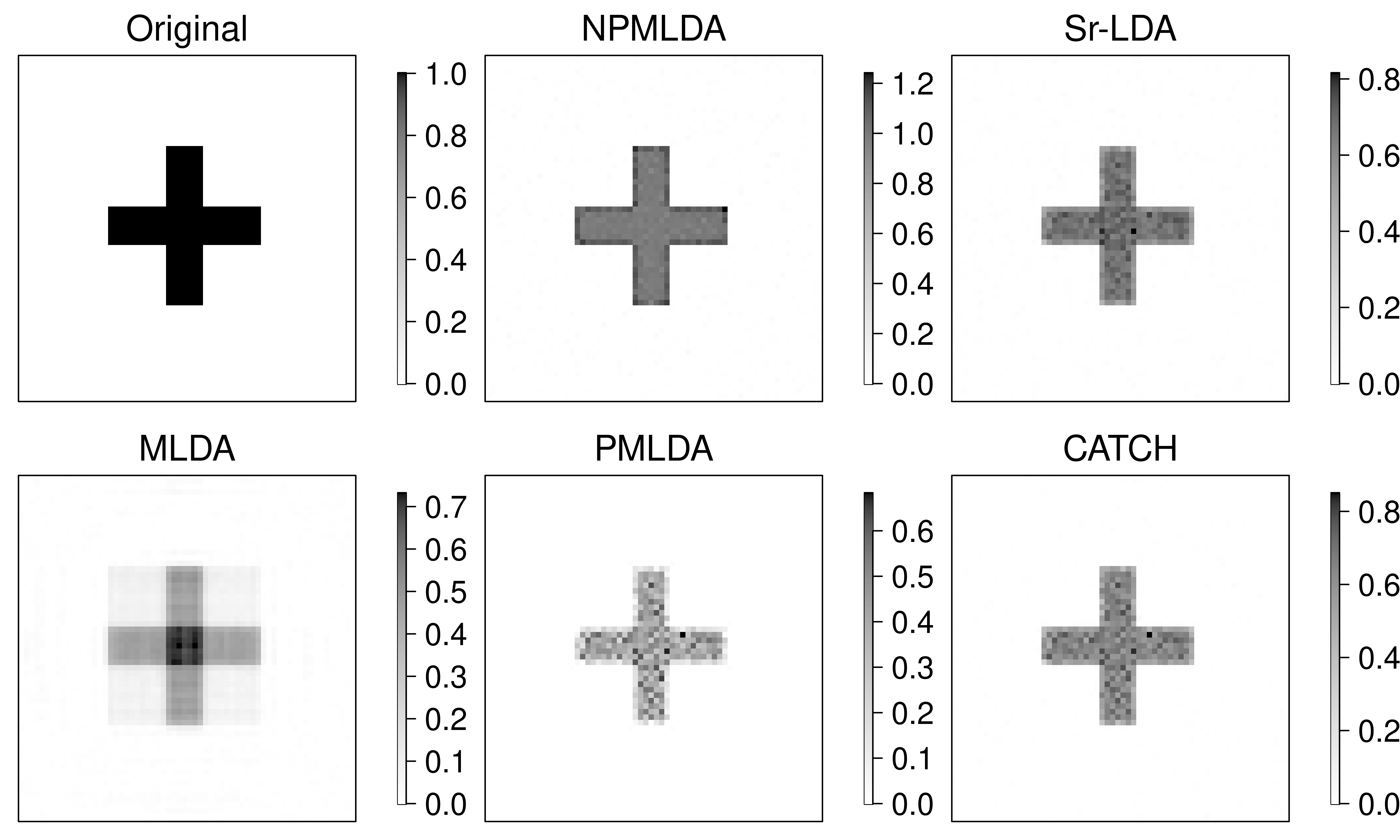}
        \caption{Averaged estimated $\mathbf{B}$ from each method under the sparse $\mathbf{B}$ setting.}
        \label{fig:est_B_sparse}
\end{figure}

\section{Real data analysis} \label{sec:da}
In this section, we demonstrate the empirical superiority of our method using real-world data. Specifically, we consider applying the electroencephalogram (EEG) dataset and magnetic resonance imaging (MRI) dataset to our method. 

\subsection{Electroencephalogram (EEG) data analysis} \label{sec:eeg}
The EEG dataset includes numerical data recorded from 64 electrodes positioned on the subjects' scalps, sampled at a frequency of 256 Hz (3.9-msec epoch) over a duration of 1 second. It is publicly available at \url{https://archive.ics.uci.edu/dataset/121/eeg+database}. There are a total of 122 subjects (77 alcoholics and 45 controls), each of whom completed 120 trials. As part of the preprocessing, referring to \cite{Park_2022} and \cite{hung_2019}, we split the 256 time points into 16 partitions, and utilize the median value of each partition as its representative value. 
Then, we compute the element-wise average of the 16-by-64 matrix data for each individual. The misclassification rate of each model is calculated using Leave-One-Out Cross-Validation (LOOCV). The results can be found in Table \ref{tab:EEG}. It is clear that our method achieves the lowest misclassification rate among all classifiers.

\begin{table} [h]
    \centering
    \begin{tabular}{lccccc}
    \toprule
    \textbf{Classifier} & \textbf{NPMLDA} & \textbf{MLDA} & \textbf{PMLDA} & \textbf{CATCH} & \textbf{Sr-LDA} \\
    \midrule
    Misclassification Rate & 0.1639 & 0.2131 & 0.1967 & 0.2049 & 0.1803 \\
    \bottomrule
    \end{tabular}
    \caption{Misclassification rates computed by LOOCV on the EEG dataset.}
    \label{tab:EEG}
\end{table}
\subsection{Magnetic resonance imaging (MRI) data analysis}
{The MRI dataset includes brain MRI images related to Alzheimer's disease (AD). It can be accessed at \url{https://github.com/Jacobheyman702/Alzheimer_Image_classifier-}. All observations are labeled according to four stages of AD symptom progression: Non Demented, Very Mild Demented, Mild Demented, and Moderate Demented. Among the possible combinations, we select two pairs for binary classification. Section \ref{sec:MRI1} covers the binary classification problem of distinguishing the Non Demented group from the Very Mild Demented group, and Section \ref{sec:MRI2} focuses on distinguishing between the Very Mild Demented group and the Mild Demented group.}

\subsubsection{Non Demented vs. Very Mild Demented} \label{sec:MRI1}
{We begin by analyzing the classification task between the \textit{Non Demented} and \textit{Very Mild Demented} groups.}
To construct a high-dimensional low sample size (HDLSS) setting, we randomly choose 50 subjects from each group; that is, 50 Non Demented and 50 Very Mild Demented subjects are used for this analysis. For preprocessing, we first rescale the 208-by-176 images to 92-by-78, and then crop them to roughly retain the brain regions while eliminating most of the uninformative black background. Subsequently, we conduct min-max scaling for each image. The Figure \ref{fig:MRI1} visualizes the group-wise averages of the preprocessed images. The performance of the models is assessed through Leave-One-Out Cross-Validation (LOOCV), and the corresponding results are presented in Table \ref{tab:MRI1}. 
{Notably, the proposed NPMLDA method achieves a zero misclassification rate, outperforming all other competing methods.}
}
\begin{figure}[H]
    \centering
    \includegraphics[width=0.7\linewidth]{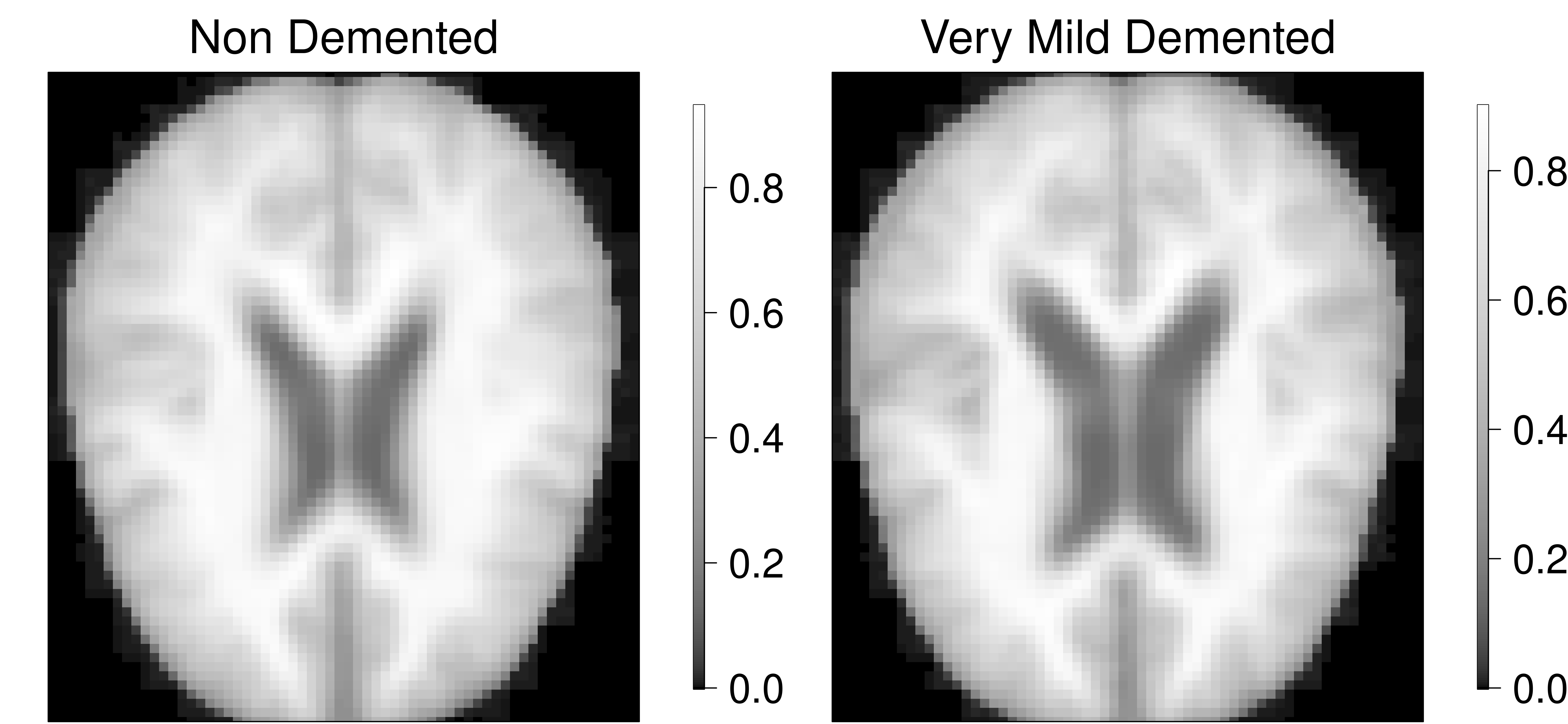}
    \caption{Averaged preprocessed brain images for Non Demented and Very Mild Demented groups.}
    \label{fig:MRI1}
\end{figure}
\begin{table}[H]
    \centering
    \begin{tabular}{lccccc}
    \toprule
    \textbf{Classifier} & \textbf{NPMLDA} & \textbf{MLDA} & \textbf{PMLDA} & \textbf{CATCH} & \textbf{Sr-LDA} \\
    \midrule
    Misclassification Rate & 0.0000 & 0.1200 & 0.0400 & 0.1600 & 0.0600 \\
    \bottomrule
    \end{tabular}
    \caption{Misclassification rates computed by LOOCV on the MRI dataset (Non Demented vs. Very Mild Demented).}
    \label{tab:MRI1}
\end{table}
\subsubsection{Very Mild Demented vs. Mild Demented} \label{sec:MRI2}
\begin{figure}[H]
            \centering
            \includegraphics[width=0.7\linewidth]{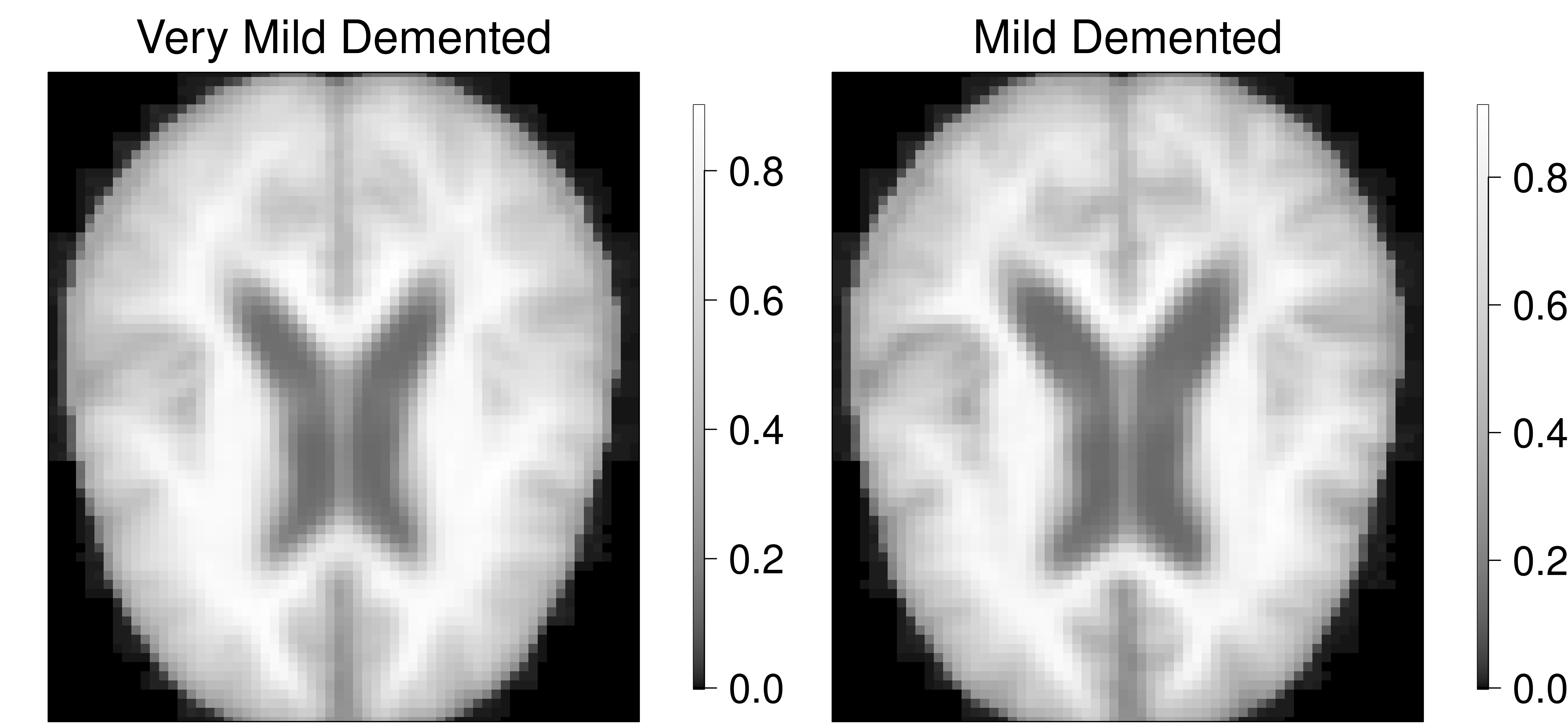}
            \caption{Averaged preprocessed brain images for Very Mild Demented and Mild Demented groups.}
            \label{fig:MRI2}
\end{figure}

{
{We next consider the more challenging task of classifying the \textit{Very Mild Demented} and \textit{Mild Demented} groups.} The same preprocessing procedures are performed as described in Section \ref{sec:MRI1}, and Figure \ref{fig:MRI2} presents the group-wise averaged preprocessed images. The results are shown in Table \ref{tab:MRI2}. Once again, NPMLDA provides the lowest misclassification rate.}

\begin{table}[H]
    \centering
    \begin{tabular}{lccccc}
    \toprule
    \textbf{Classifier} & \textbf{NPMLDA} & \textbf{MLDA} & \textbf{PMLDA} & \textbf{CATCH} & \textbf{Sr-LDA} \\
    \midrule
    Misclassification Rate & 0.1800 & 0.2500 & 0.2600 & 0.2300 & 0.1900 \\
    \bottomrule
    \end{tabular}
    \caption{Misclassification rates computed by LOOCV on the MRI dataset (Very Mild Demented vs. Mild Demented).}
    \label{tab:MRI2}
\end{table}

\subsubsection{Non Demented vs. Very Mild Demented vs. Mild Demented} \label{sec:MRI3}

{Lastly, we address the multi-class problem of classifying the \textit{Non Demented},  \textit{Very Mild Demented} and \textit{Mild Demented} groups. The same dataset as in Section \ref{sec:MRI1} and \ref{sec:MRI2} are used, and they are preprocessed in the same manner. Since MLDA, and Sr-LDA are designed for binary classification, this section only presents the results of NPMLDA, PMLDA, and CATCH. The Table \ref{tab:MRI3} contains the results.}

\begin{table}[H]
    \centering
    \begin{tabular}{lccccc}
    \toprule
    \textbf{Classifier} & \textbf{NPMLDA} & \textbf{MLDA} & \textbf{PMLDA} & \textbf{CATCH} & \textbf{Sr-LDA} \\
    \midrule
    Misclassification Rate & 0.1067 & - & 0.1933 & 0.2267 & - \\
    \bottomrule
    \end{tabular}
    \caption{Misclassification rates computed by LOOCV on the MRI dataset (Non Demented vs. Very Mild Demented vs. Mild Demented).}
    \label{tab:MRI3}
\end{table}

{Additionally, based on the datasets in Sections \ref{sec:eeg}, \ref{sec:MRI1}, \ref{sec:MRI2}, and \ref{sec:MRI3}, we also performed k-fold cross-validation to assess the variability of the misclassification rates. The results are provided in the Supplementary Material.}

\section{Conclusion} \label{sec:conclusion}
{In this paper, we proposed a novel, assumption-relaxed classification framework for matrix-valued data. Our method demonstrated remarkable flexibility across a wide range of data structures.}
{Through extensive simulation studies and real-world applications—including the analysis of EEG recordings and MRI data from Alzheimer disease research—we provided empirical evidence supporting the effectiveness of the proposed approach. In the simulation studies, which considered a wide range of data-generating scenarios, the superiority of our method was particularly evident in its ability to accurately recover the underlying coefficient matrix $\mathbf{B}$ of the discriminant function. This structural recovery underscores the model's capacity to capture essential discriminative information embedded in matrix-valued data. 
Consistent with the simulation results, the low misclassification rates observed in the real data analyses further 
suggest that the proposed nonparametric method can offer stable and reliable performance.}

{While our proposed method provides the notable contributions to matrix-valued data analysis, several limitations need to be discussed. 
First, the matrix normal distribution assumption may not be appropriate in certain situations. As this assumption implies the Kronecker product structure of covariance matrices in the two-dimensional matrix-valued setting, our method can be affected when the true covariance cannot be decomposed into the Kronecker product form. Therefore, it is necessary to examine its robustness when this assumption is violated, as well as to explore ways to relax the normality assumption. Second, the estimation of the scaled mean parameters in Section \ref{eqn:subsec:parameter_estimation} depends on the estimation of the covariance matrices. Thus, readers should carefully choose the appropriate covariance estimation method according to their analysis context.}

Apart from the points discussed above, we plan to extend our framework to accommodate high-dimensional tensor data. By generalizing the current methodology to multiway arrays, we aim to develop a flexible and scalable classification technique capable of capturing the intricate dependencies within tensor-valued data. In addition, we intend to examine the rigorous theoretical properties of our method, including asymptotic consistency and convergence rates to the Bayes optimal classifier. We plan to establish that the misclassification error approaches zero under suitable regularity conditions and derive finite-sample bounds for the classification performance. Additional theoretical investigations will focus on the optimality properties of our nonparametric approach and its robustness across different data-generating mechanisms. These future directions would provide our method with enhanced flexibility and theoretical guarantees, further extending its applicability to modern datasets encountered in a variety of complex scientific and engineering problems.

\section*{Code Availability}
{The R codes used to implement the NPMLDA method, including a simple example for one of the same settings as those in Section \ref{sec:sim}, are available at \url{https://github.com/SeungyeonOh-1999/Nonparametric-matrix-linear-discriminant-analysis}.}

\section*{Acknowledgement}
Seongoh Park and Hoyoung Park were supported by the government of the Republic of Korea (MSIT) and the National Research Foundation of Korea (RS-2024-00338876), and Hoyoung Park was supported by the National Research Foundation of Korea (NRF) grant funded by the Korea government(MSIT) (No. RS-2023-00212502)

\bibliographystyle{apalike}  
\bibliography{Refs} 

\newpage
\section*{Supplementary Material}

\renewcommand{\thefigure}{S\arabic{figure}}
\renewcommand{\thetable}{S\arabic{table}}
\setcounter{figure}{0}
\setcounter{table}{0}

{All the additional scenarios follow the same settings as in Section \ref{sec:sim}, except for the size of the cross pattern of $\mathbf{B}$.}
\begin{figure} [H]
    \centering
    \includegraphics[width=1\linewidth]{figures/supp/B.png}
    \caption{Examples of $\mathbf{B}$ at $\theta = 1$ by cross pattern size.}
    \label{fig:supp_B}
\end{figure}
\subsection*{Small cross}

\begin{figure}[H]
    \centering
    \includegraphics[width=0.85\linewidth]{figures/supp/small/s_dense_model1.png}
    \caption{Misclassification rates in Model 1 under the dense $\mathbf{B}$ setting with the small cross.}
    \label{fig:supp_s_dense_M1}
\end{figure}

\begin{figure}[H]
    \centering
    \includegraphics[width=0.85\linewidth]{figures/supp/small/s_sparse_model1.png}
    \caption{Misclassification rates in Model 1 under the sparse $\mathbf{B}$ setting with the small cross.}
    \label{fig:supp_s_sparse_M1}
\end{figure}


\begin{figure}[H]
    \centering
    \includegraphics[width=0.85\linewidth]{figures/supp/small/s_dense_model2.png}
    \caption{Misclassification rates in Model 2 under the dense $\mathbf{B}$ setting with the small cross.}
    \label{fig:supp_s_dense_M2}
\end{figure}

\begin{figure}[H]
    \centering
    \includegraphics[width=0.85\linewidth]{figures/supp/small/s_sparse_model2.png}
    \caption{Misclassification rates in Model 2 under the sparse $\mathbf{B}$ setting with the small cross.}
    \label{fig:supp_s_sparse_M2}
\end{figure}


\begin{figure}[H]
    \centering
    \includegraphics[width=0.85\linewidth]{figures/supp/small/s_dense_model3.png}
    \caption{Misclassification rates in Model 3 under the dense $\mathbf{B}$ setting with the small cross.}
    \label{fig:supp_s_dense_M3}
\end{figure}

\begin{figure}[H]
    \centering
    \includegraphics[width=0.85\linewidth]{figures/supp/small/s_sparse_model3.png}
    \caption{Misclassification rates in Model 3 under the sparse $\mathbf{B}$ setting with the small cross.}
    \label{fig:supp_s_sparse_M3}
\end{figure}

\subsection*{Large cross}

\begin{figure}[H]
    \centering
    \includegraphics[width=0.85\linewidth]{figures/supp/large/l_dense_model1.png}
    \caption{Misclassification rates in Model 1 under the dense $\mathbf{B}$ setting with the large cross.}
    \label{fig:supp_l_dense_M1}
\end{figure}

\begin{figure}[H]
    \centering
    \includegraphics[width=0.85\linewidth]{figures/supp/large/l_sparse_model1.png}
    \caption{Misclassification rates in Model 1 under the sparse $\mathbf{B}$ setting with the large cross.}
    \label{fig:supp_l_sparse_M1}
\end{figure}


\begin{figure}[H]
    \centering
    \includegraphics[width=0.85\linewidth]{figures/supp/large/l_dense_model2.png}
    \caption{Misclassification rates in Model 2 under the dense $\mathbf{B}$ setting with the large cross.}
    \label{fig:supp_l_dense_M2}
\end{figure}

\begin{figure}[H]
    \centering
    \includegraphics[width=0.85\linewidth]{figures/supp/large/l_sparse_model2.png}
    \caption{Misclassification rates in Model 2 under the sparse $\mathbf{B}$ setting with the large cross.}
    \label{fig:supp_l_sparse_M2}
\end{figure}

\begin{figure}[H]
    \centering
    \includegraphics[width=0.85\linewidth]{figures/supp/large/l_dense_model3.png}
    \caption{Misclassification rates in Model 3 under the dense $\mathbf{B}$ setting with the large cross.}
    \label{fig:supp_l_dense_M3}
\end{figure}

\begin{figure}[H]
    \centering
    \includegraphics[width=0.85\linewidth]{figures/supp/large/l_sparse_model3.png}
    \caption{Misclassification rates in Model 3 under the sparse $\mathbf{B}$ setting with the large cross.}
    \label{fig:supp_l_sparse_M3}
\end{figure}

\subsection*{Real data analysis results using 10-fold cross-validation.}
\begin{table} [H]
    \centering
    \begin{tabular}{l c c c c}
    \toprule
    \textbf{Dataset} & \textbf{EEG} & \textbf{MRI 1} & \textbf{MRI 2} & \textbf{MRI 3} \\
    \midrule
    NPMLDA & 0.1724 (0.1076) & 0.0000 (0.0000) & 0.1632  (0.1258) & 0.1066 (0.0573) \\
    MLDA   & 0.2346 (0.1381) & 0.0915 (0.0744) & 0.1834 (0.0794) & - \\
    PMLDA  & 0.2692 (0.1535) & 0.0837 (0.1359) & 0.2236 (0.0890) & 0.2072 (0.1183) \\
    CATCH  & 0.2199 (0.1134) & 0.1848 (0.1782) & 0.2609 (0.0970) & 0.2048 (0.1483) \\
    Sr-LDA & 0.2442 (0.1304) & 0.0677 (0.0881) & 0.1905 (0.0330) & - \\
    \bottomrule
    \end{tabular}
    \caption{Misclassification rates computed by 10-fold cross-validation with associated standard errors: The EEG column reports the results for the EEG dataset used in Section \ref{sec:eeg}. The three MRI columns are associated with the MRI datasets used in Sections \ref{sec:MRI1}, \ref{sec:MRI2}, and \ref{sec:MRI3}, respectively.}
    \label{tab:kfolds}
\end{table}


\end{document}